\documentclass[preprint,aps,showpacs,eqsecnum,nofootinbib,12pt]{article}
\usepackage{amsfonts}
\usepackage{array}
\usepackage{color}
\usepackage{amsmath}
\usepackage{amsxtra}
\usepackage{amstext}
\usepackage{amssymb}
\usepackage{latexsym}

\usepackage{epsfig}
\usepackage{amsfonts}
\usepackage{mathrsfs}
\usepackage{graphicx}

\definecolor{dyellow}{rgb}{1.,0.8,.0}
\definecolor{marble}{rgb}{.1,.1,.7}
\definecolor{cyan}{rgb}{.0,.6,.6}
\definecolor{magenta}{rgb}{0.6,0.0,0.6}
\definecolor{brown}{rgb}{0.6,0.2,0.}
\definecolor{markable}{rgb}{.0,.0,0.5}
\definecolor{darken}{rgb}{0.75,0.0,0.0}
\definecolor{orange}{rgb}{1.,.6,.0}
\definecolor{derange}{rgb}{0.8,.4,.0}
\definecolor{carageen}{rgb}{0.0,0.6,0.0}
\definecolor{purple}{rgb}{.4,.0,.4}

\def\red{\color{red}}
\def\blue{\color{blue}}

\def\green{\color{green}}

\newcommand{\omits}[1]{}

\newcommand{\A}{$\cal A$}
\def\disp{\displaystyle}
\def\SL{\SL(2,\mathbb{C})}

\def\H1{H_{x_1}}

\def\dfrac{\displaystyle\frac}
\def\dsum{\displaystyle\sum}

\def\ba{\begin{array}}
\def\ea{\end{array}}
\def\bd{\begin{displaymath}}
\def\ed{\end{displaymath}}
\def\be{\begin{equation}}
\def\ee{\end{equation}}
\def\ben{\begin{eqnarray*}}
\def\een{\end{eqnarray*}}
\def\bes{\begin{eqnarray}}
\def\ees{\end{eqnarray}}

\begin{document}

\begin{center}
 {\large\bf DIFFERENCE DISCRETE CONNECTION  AND \\[2mm]CURVATURE ON CUBIC
 LATTICE} \vskip3mm

\end{center}\vskip4mm

\vskip5pt \centerline{Ke Wu, \quad Wei-Zhong Zhao } \vskip5pt
\centerline{ Department of Mathematics, Capital Normal University,  Beijing 100037, China} 
 \vskip15pt \centerline{Han-Ying Guo} \vskip5pt
\centerline{ Institute of Theoretical Physics, Chinese Academy of
Sciences, Beijing 100080, China} \vskip30pt

\begin{center}
\begin{minipage}{130mm}
\vskip 2.5 cm
\begin{center}
\bf{Abstract}
\end{center} In a way
similar to the continuous case formally, we define in different
but equivalent manners  the difference discrete connection and
curvature on discrete vector bundle over  the regular lattice as
base space. We deal with the difference operators as the discrete
counterparts of the
 derivatives based upon the differential
calculus on
 the lattice. One of the definitions can be extended to the case over the random lattice.
 \omits{and its relation with the recent paper by Leok et al are
discussed. }We also discuss the relation between our approach and
the lattice gauge theory and apply to the discrete integrable
systems.\vskip 27mm

\end{minipage}
\end{center}


\newpage
\section{Introduction}

 The discrete systems play very important roles in various fields, so they
 are widely studied in different branches.
 As  one of the successful ways to deal with the quantum gauge field
theory non-perturbatively, the lattice gauge theory in high energy
physics has opened up a new direction in both physics and
mathematics in order to deal with the gauge potentials as the
connections in a discrete manner.
 For the integrable systems, there are a kind of discrete integrable ones
 as
 the discrete counterparts of the continuous systems by
 means of the integrable discretization method.
For the structure preserving algorithms or the geometric algorithms
in computational mathematics, the continuous systems are discretized
in such a way that some of the important properties such as the
symplectic or multisymplectic structure, or the symmetry of the
systems are required to be preserved in a discrete manner,
respectively. There are many other important discrete systems and
some of them are even without the proper or unique continuous
limits. In this paper, we focus on how to get the discrete
counterparts in a systematic manner for such a kind of continuous
systems that there are important properties like the gauge
potentials as connections, the symplectic or multisymplectic
structures, the Lax pairs, the symmetries and so on.

A simple and direct method to get a  discrete counterpart for a
given continuous system such as an ODE or a PDE is to discrete the
independent variable(s) and let the dependent variables become
discrete accordingly without specially chosen.
 However, in
most cases all important properties of the continuous system may
be lost and the behaviors of the discrete system are even hard to
be controlled. For constructing mostly quarried ones in all
possible discretizations of the corresponding continuous systems,
there is a working guide line or a structure-preserving criterion.
 Namely, it is important to look forward to those
discrete systems that preserve as much as possible the intrinsically
important properties of the continuous system (see, e.g.
\cite{fk84}, \cite{fk95}, \cite{ GLWW1}, \cite{ GLWW2}, \cite{
GLWW3}, \cite{GW}).

In the course of discretization, only a few of the most important
 properties, or ``structure" could be maintained.
Thus, it is needed to select these ``structures",  to find  their
discrete counterparts and to know how to preserve them discretely
with the lowest price has to pay. For example, there are two classes
of conservation laws in canonical conservative mechanics. The first
class is  of phase-area conservation laws characterized by the
symplectic preserving property.  Another class is related to energy
and all first integrals of the canonical equations. Thus, it is
needed to know if it is possible to establish such a kind of
discrete systems that they not only discretely preserve the
``structure", such as the symplectic structure, but also  the energy
conservation discretely. And is it possible to get these discrete
systems by a discrete variational principle?

In fact, as far as the discrete variation for the discrete mechanics
are concerned, there are different approaches. In usual approach
(see, e.g. \cite{TD1}, \cite{TD2}, \cite{TD3}, \cite{Veselov1},
\cite{Veselov2}), only the discrete dependent variables are taken as
the independent variational variables, while their differences are
not. However, in the discrete variational principle proposed by two
of the present authors and their collaborators recently
\cite{GLWW1}, \cite{GLWW2}, \cite{GLWW3}, \cite{GW}, the differences
of the dependent variables as the discrete counterparts of
derivatives are taken as the independent variational variables
together with the discrete dependent variables themselves. Actually,
this is just the discrete analogue of the variational principle for
the continuous mechanics{\red,} where the derivatives of the
dependent variables are dealt with  as the independent variational
variables first. Thus, the difference discrete Legendre
transformation can be made and the method can be applied to either
discrete Lagrangian mechanics or its Hamiltonian counterpart via the
difference discrete Legendre transformation. The approach has been
applied to the symplectic and multisymplectic algorithms in the both
Lagrangian and Hamiltonian formalism. It has been also  generalized
to the case of variable steps in order to preserve the discrete
energy in addition to the symplectic or the multisymplectic
structure\cite{GW}, \cite{TD1}, \cite{TD2}, \cite{TD3}, \cite{
LGLW}.

For the lattice gauge theory as the discrete counterpart of the
gauge theory in continuous spacetime, the discrete gauge potentials,
field strength and the action had been introduced in a manner almost
completely different from, as least  apparently, the ones in
ordinary gauge fields or in the connection theory \cite{DFNC1},
\cite{DFNC2} in fibre bundle. Although the discrete connection
theory also has its own right (see, e.g. \cite{Nov,LMW}) as an
application of the non-commutative geometry \cite{Con}, how to
introduce the discrete gauge potential in the lattice gauge theory
as a kind of discrete connection is an important and interesting
problem. Two of us with their collaborator had considered this issue
in \cite{GWZ} very briefly in a way  different from other relevant
proposals (see, e.g. \cite{Sit}, \cite{Dim94}, \cite{Dim941},
\cite{Dim942},\cite{Dim943}).

In this  paper, we  study the discrete connection and curvature on
the regular lattice further in a  way similar to the connection and
curvature on vector bundle.  We define them in different but
equivalent manners, find their gauge transformation properties, the
Chern class in the Abelian case and relevant issues. We  show that
the discrete connection and curvature introduce here are completely
equivalent to the ones in the lattice gauge theory on the regular
lattice. We also apply these issues to the discrete integrable
systems and show that their discrete Lax pairs and the
discrete-curvature free conditions are certainly  similar to  their
continuous cases.

One of the key points in our approach is still to regard the
difference operators acting on the functions space over the regular
lattice as a kind of independent geometric objects and their dual
should be the one forms such that we can introduce the discrete
tangent bundle over the lattice and the cotangent bundle as its dual
with the basis as the difference operators and one-formes,
respectively. These are just the discrete counterparts of the
continuous cases, where the derivatives and the one-formes as their
dual play the roles as the basis of the tangent bundle and the
cotangent bundle, respectively.  In order to do so, the
non-commutative differential calculus on the function space over the
lattice \cite{GWZ} has to be employed. Similarly, the discrete
vector bundle over the regular lattice can also be set up. In the
connection theory a la Cartan, the exterior differential of the
basis of a vector space at a point should be expanded in terms of
the basis and the expanding coefficients are just the coefficients
of the connection on the vector space. Since all counterparts of the
basis, exterior differential and so on are equipped in the discrete
vector bundle over the lattice, the discrete connection can also be
introduced in a way similar to the one a la Cartan. This is our
simple way to introduce the discrete connection. In continuous
cases, there are several equivalence definitions for the connection
and curvature. Similarly, we also introduce some of the equivalent
definitions for the discrete connection and curvature on the
lattice. It should be noted that the definition of the discrete
connection in terms of the decomposition of the tangent space of the
discrete vector bundle may be written in the form without
differences. Thus, it can be generalized to the cases over the
random lattices.

 The paper is organized as follows. In
order to show the background and necessary preparation,  we briefly
recall the discrete mechanics and the non-commutative differential
calculus on hypercubic lattice of high dimension in section 2 and
section 3, respectively. In section 4, we discuss the discrete
connection, curvature, Chern class and their gauge transformation
properties. In section 5, we generalize one of the definitions for
the discrete connection to the case of random lattice. Some
applications to the lattice gauge theory and discrete integrable
systems  are given in section 6. We end with some remarks and
discussions.

 \section{Difference Discrete Mechanics}\label{DDM}
\hskip6mm  In order to introduce the structure of discrete bundle,
it is useful to review the formulism in  the Lagrangian mechanics
and  the discrete mechanics  as its discrete counterpart.

Let \omits{\blue us consider the cases that there is a translation
invariance on the time, i.e. }$t \in T $ be the time, $M$ an
$n$-dimensional configuration space as a vector space for
simplicity. A particle moving on the configuration space is
denoted in terms of its generalized coordinates as $q^i(t)\in M$
and its generalized velocities $\dot q^i(t)=dq^i(t)/dt$ as an
element in tangent bundle $TM$ of $M$. The space of all the
possible path of particle moving in configuration space is an
infinite dimensional space. \omits{Consider a fibre bundle $E(T,
Q, \pi)$ with projection $\pi: E \rightarrow T$ on $T$,
${\pi}^{-1}: t\rightarrow Q_t$ isomorphic to $M$ is the fibre on
$t \in T$. Denote $\Gamma(E)$ the sections on $E$, $TE$ the
tangent bundle of $E$, $T_v E\subset TE$ the vertical sub-bundle
of $TE$, etc..} The Lagrangian of a system is a functional defined
on this space and denoted as
 $L(q^i(t), \dot q^i(t))$,
$i=1,\cdots ,n$.  For simplicity, the Lagrangian in our discussion
is of the first order and independent of $t$. The action
functional is
\begin{equation}
{S}([q^i(t)]; t_1, t_2)=\int_{t_1}^{t_2}dt {L}(q^i(t), \dot
q^i(t)). \label{actn}
\end{equation}
Here 
$q^i(t)$ describes a curve ${C}_a^b$ with ending points $a$ and
$b$, $t_a=t_1, t_b=t_2$, along which the motion of the system  may
takes place.

For the difference discrete Lagrangian mechanics,  let us consider
the case that ``time" $t$ is difference discretized
\begin{equation}\label{td}
t\in T  \rightarrow t_k\in  {T_D}=\{ (t_k , t_{k+1}=t_k+\Delta
t_k=t_k+h, \quad k \in Z)\}
\end{equation}
and the step-lengths $\Delta t_k=h$ are equal to each other for
simplicity,
 while the $n$-dimensional
configuration space $M_k$ at each moment $t_k, k \in Z$, is still
continuous and smooth enough.

Let $N$ be the set of all nodes on ${T_D}$ with index set
$Ind({N})=Z$, ${M}=\bigcup_{k \in Z} M_k$ the configuration space
on $T_D$ that is 
at least pierce wisely smooth enough. At the moment $t_k$, ${\cal
N}_k$  be the set of nodes neighboring to $t_k$.
 Let ${I}_k $ the index set of nodes of ${\cal N}_k$ including $t_k$.
   The coordinates of $M_k$
 are   denoted  by $q^i({t_k})=q^{i (k)}, i=1, \cdots,
n$.  $T(M_k)$ the tangent bundle of $M_k$ in the sense that
difference at $t_k$ is its base, $T^*(M_k)$ its dual. Let ${\cal
M}_k=\bigcup_{{l} \in {I}_k} M_l $ be the union of configuration
spaces $M_l$ at $t_l, {l} \in {I}_k$ on ${\cal N}_k$, $T{\cal
M}_k=\bigcup_{{l}\in {I}_k}TM_l $ the union of tangent bundles on
${M}_k$, $F(TM_k)$ and $F(T{\cal M}_k)$ the  functional spaces on
each of them respectively, etc.. \omits{\red The notation ${\cal
M}_k$ used here is not only to indicate the discrete velocities are
located on the nearest points, but also the generalized coordinates
are sometimes taking value on the nearest points, like the middle
point scheme in symplectic algorithm.} Sometime, it is also needed
to include the links, plaquette and so on as well as the dual
lattice, like in the lattice gauge theory,  the mid-point scheme in
the symplectic algorithm and so on. In these cases, the  notations
and conceptions introduced here should be generalized accordingly.

\omits{\blue Remarks: How to deal with links, plaquette and so on
at the lattice? In the Euler mid-point scheme, $g^{k+1/2}$ has to
be used. How to explain it?}

 The above consideration should also make sense for the
vector bundle over either the 1-dimensional lattice $T_D$ or the
higher dimensional lattice as a discrete base manifold. In the
difference variational approach and the definition of the
difference discrete connection, these notions may be used.

Now the discrete Lagrangian ${{L}_D}^{(k)}$  on $F(T({M}_k ))$ reads
\omits{.  For simplicity, the Lagrangian is of the first order of
differences}
\begin{equation}\label{lmd}
{{L}_D}^{(k)}={L}_D( q^{i (k)}, {\Delta_{k} q}^{i (k)}),
\end{equation}
with the difference ${\Delta_{k} q}^{i (k)}$ of $q^{i (k)}$ at
$t_k$ defined by
\begin{equation}\label{dfc}
{\Delta_{k} q}^{i (k)}:=\frac{q^{i (k+1)}-q^{i
(k)}}{t_{k+1}-t_{k}}=\frac{1}{h}(q^{i (k+1)}-q^{i (k)}).
\end{equation}
 The discrete action of the system is given by
\begin{equation}
{S}_D=\sum_{k \in Z} h\cdot {{L}_D}^{(k)}(q^{i (k)}, {\Delta_{k}
q}^{i (k)}).
\end{equation}

The discrete  variation for $q^{i (k)}=q^{i}(t_k)$
 should be defined as
\begin{equation}\label{tvdq}
\delta q^{i (k)}:=q'^i(t_k)-q^i (t_k).%
\end{equation}%
And the discrete variations for ${\Delta_k q}^{i (k)}$ are given
by \omits{defined as
\begin{equation}\label{tvDq}
\delta \Delta_k q^{i (k)}:=\frac{1}{h}(q'^i(t_{k+1})-q'^i(t_{k}))-
\frac{1}{h}(q^i(t_{k+1})-q^i(t_{k})).
\end{equation}%
 Due to the definition of the difference with
variable time step-length (\ref{dfc}) and the Leibnitz law for it
\begin{equation}\label{lbnz}
\Delta_k (f^{(k)}g^{(k)})=(\Delta_k f^{(k)})g^{(k)}+
Ef^{(k)}(\Delta_k g^{(k)}),
\end{equation}
where $E$ is the shift operator defined as
\begin{equation}\label{shft}
Ef^{(k)}=f^{(k+1)},\qquad E^{-1}f^{(k)}=f^{(k-1)},
\end{equation}
 it follows}
\begin{equation}\label{tvDq3}
\delta \Delta_k q^{i (k)}=\Delta_k \delta q^{i (k)}.
\end{equation}
Thus, the  variations of the discrete Lagrangian can be calculated
\begin{equation}\label{tvdL}
\delta{{L}_D}^{(k)} =[L_{q^{i (k) }}]\delta q^{i (k)} +\Delta_k (
{p_i}^{(k+1)} \delta q^{i (k)}), \end{equation} where $[L_{q^{i
(k) }}]$ is the discrete Euler-Lagrange operator
\begin{equation}\label{eloD}
[L_{q^{i (k) }}]:=\frac{\partial{{ L}_D}^{(k)}}{\partial q^{i
(k)}}-\Delta \big ( \frac{\partial{{ L}_D}^{(k-1)}}{\partial
\Delta q^{i (k-1)}}\big),
\end{equation}
and ${p_i}^{(k)}$ the  discrete canonical conjugate momenta
\begin{equation}\label{mmntad}
{p_i}^{(k)}:=\frac{\partial{{L}_D}^{(k-1)}}{\partial \Delta q^{i
(k-1)}}.
\end{equation}
And the  variation of the discrete action 
 can be
written as
\begin{equation}\label{tvds2}\omits{\begin{array}{rcl}
\delta_t{S}_D&=&\delta_v {S}_D+\delta_h {S}_D,\\}
\delta{S}_D=\sum_k h
[L_{q^{i (k) }}]\delta_v q^{i
(k)}+\Delta ({p_{i}}^{(k+1)}
\delta_vq^{i (k)}).\omits{\\
\delta_h {S}_D&=&\sum_k (t_{k+1}-t_k)\{[L_{q^{i (k) }}]\delta_h
q^{i (k)}+(\Delta{{H}_D}^{(k-1)} +\frac{\partial{{
L}_D}^{(k)}}{\partial t_k}) \delta t_k \\&+&\Delta
({p_{i}}^{(k+1)} \delta_h q^{i (k)}- {{H}_D}^{(k-1)}\delta t_k)\}.
\end{array}}\end{equation}

The variational principle requires $\delta {S}_D=0$, so it follows
the discrete Euler-Lagrange equations for $q^{i (k)}$'s
\begin{equation}\label{eleqd}
\frac{\partial{{L}_D}^{(k)}}{\partial q^{i (k)}}-\Delta (
\frac{\partial{{L}_D}^{(k-1)}}{\partial \Delta q^{i
(k-1)}})=0.%
\end{equation}

In order to transfer to the discrete  Hamiltonian formalism, it is
needed to introduce the discrete canonical conjugate momenta
according to the equation (\ref{mmntad}) and express the discrete
Lagrangian by the discrete Hamiltonian via  Legentre
transformation
\begin{equation}\label{lgdrd}
{{H}_D}^{(k)}:={p_{i}}^{(k+1)}\Delta_t q^{i
(k)}-{{L}_D}^{(k)}.%
\end{equation}%
 Thus, the discrete action can be
expressed as
\begin{equation}\label{actndh}
{S}_D
=\sum_{k}h\cdot 
({p_{i}}^{(k+1)}\Delta_t q^{i (k)}-{{
H}_D}^{(k)}).
\end{equation}
Now, Hamilton's 
principle requires $\delta_v {S}_D=0$, it follows the discrete
canonical equations for ${p_i}^{(k)}$'s and $q^{i (k)}$'s
\begin{equation}\label{cneqd}
\Delta q^{i (k)}= \frac{\partial{{H}_D}^{(k)}}{\partial
{p_i}^{(k+1)}},\quad \Delta {p_i}^{(k)}=-\frac{\partial{{
H}_D}^{(k)}}{\partial {q}^{i (k)}}.%
\end{equation}

As was mentioned,  the advantages of the difference discrete
variational principle are based on keeping the difference operator
as a discrete derivative operator. It is also clear that this
approach can be applied to the field theory as well and
generalized to the total discrete variation with variable
step-lengths \cite{GW, LGLW}. Actually, this key point will also
play a central role in our proposal to the discrete connection and
curvature.

 In the usual discrete variation, however,  the  $Q\times
Q$  is used to indicate the vector field on the discrete space and
the difference has not been dealt with as independent variables
(see, e.g. \cite{TD1}, \cite{TD2}, \cite{TD3}, \cite{DiscVar},
\cite{DiscVar1}, \cite{DiscVar2}). The corresponding discrete action
is
\begin{equation}
S=\sum_{k=0}^{n-1} h\cdot \mathbb{L}(q_{k},q_{k+1}),
\end{equation}
 where the Lagrangian $ \mathbb{L}(q_{k},q_{k+1})$ is the
functional on $Q\times Q$. This is also the central idea of the
resent proposal to the disconnection in  \cite{LMW}. Namely, using
the tensor product $Q\times Q$ to study discrete tangent space of
$Q$. In other words,  the tangent vector $\dot q(t)$ at $t_k$ is
represented by a pair of nodes $(q_k, q_{k+1})$ without introducing
the difference operator. Thus, the difference discrete Legendre
transformation and discrete Hamiltonian formalism via the
transformation cannot be formulated.  In this case it is expected
that the groupoid formulism may be used so that there is possibility
to understand   some geometric meanings of discrete models
\cite{groupoid}.

\section{Difference and Differential Form on Lattice}\label{ncdc} \indent
\hskip6mm
  In this section, we  recall how to apply the differential calculus on discrete group \cite{Sit}
 to the  hypercubic lattice \cite{GWZ}. Although the result is similar to the one in
 (\cite{Dim94}, \cite{Dim941},
\cite{Dim942},\cite{Dim943})
 the key point is different.
 In our approach,  the
shift operator is regarded as the generator of a discrete Abelian
group in each direction of the high dimensional hypercubic
lattice. For simplicity, we focus on the lattice with equal
spacing $h=1$. Thus, the dimension of vector fields or
differential forms is equal to the number of the shift operators
of the lattice.  

Let $N$ and \A\ be a lattice and the algebra of complex valued
functions on $N$, respectively, define the right and left shift
operators $E_\mu, E_\mu^{-1}$ at a node $x\in N$ in the
$\mu$-direction
 by
\be\label{shift}%
E_\mu x=x+\hat \mu,\quad E_\mu^{-1}x=x-\hat\mu,
 \ee%
and introduce  a homeomorphism on the function space \A,
\be%
E_\mu f(x)=f(x+\hat \mu),\quad  E_\mu (f(x)\cdot
g(x))=E_\mu f(x)\cdot E_\mu g(x), \quad f, g \in {\cal A},%
\ee%
where  $(x-\hat\mu)$, $x$ and $(x+\hat \mu)$  are the points on $
\mathcal N_x$ and they are the nearest neighbors on
the $\mu$-direction,  the dot denotes the  multiplication in \A.%

The tangent space at the node $x$ of $
\mathcal N_x$  is 
defined as $ T{\cal N}_x :=\rm span \{ \Delta _\mu|_x,
\mu=1,\cdots, n\},$ where the  operator $\Delta_\mu$ is  defined
 on the link between $x$ and $x+\hat {\mu}$ and its action on \A\ is
the differences in $\mu$-th direction  as,
\be\label{dffrc}
\Delta_\mu f(x):= (E_\mu-id)f =f(x+\hat \mu)-f(x).%
\ee  The above difference operator is a discrete analogue of { a
bases $\partial_\nu:=\frac{\partial}{\partial x^\nu}$ for a vector
field $X=X^\nu \partial_\nu$ in the continuous case.\omits { which
is defined on the link between $x$ and $(x+\hat{\mu})$.}

The action of a difference operator $\Delta _\mu$ in
(\ref{dffrc}) on the functional space \A\
 \omits {\be%
 \Delta_\mu
f(x)=f(x+\hat{\mu})-f(x),%
\ee%
and} satisfies the deformed Leibnitz rule%
 \be\label{Lbnz}%
  \Delta_\mu
(f(x)\cdot g(x))=\Delta_\mu f(x)\cdot
g(x+\hat{\mu})+f(x)\Delta_\mu g(x).%
\ee%
 For a given node $x\in
\mathcal N_x$, all $\Delta_\mu$ form a set of bases of the tangent
space $T{\cal N}_x$.  Its dual space denoted as $T^*{\cal N}_x$ is
a space of  $1$-forms  with a set of bases
 $dx^\mu$  defined on the link, too. They  satisfy%
  \bd
 dx^\mu(\Delta_\nu)=\delta^\mu_\nu, \ed
 which is  also denoted as $\Omega^1$ and  $\Omega^0=\cal A $ like the continuous case.
  \omits{In analog to the continuous case, the
tangent bundle $T\cal N$ and its dual $T^*\cal N$ over the
discrete base space $\cal N$ may be introduced.}

Thus, the tangent bundle and its dual cotangent bundle over $N$
can be defined as \be\label{TTdual}
 T{ N}:=\bigcup_{x\in \cal
N} T{\cal N}_x \quad \mbox{and}\quad
 T^{*}{
N}:={\bigcup_{x\in { N}}}  T^{*}{ N}_x,
 \ee
  respectively.

 Let us
construct the whole differential algebra
$\Omega^*=\bigoplus\limits_{n\in Z} \Omega^n $ on $T^*{ N}$ as in
continuous case. The exterior derivative operator $d_D : \Omega^k
\rightarrow \Omega^{k+1}$ is defined as
\be d_D\omega=\sum_{\alpha}\Delta_\alpha f dx^\alpha\wedge
dx^{\mu_1}\wedge\cdots\wedge dx^{\mu_k}\in \Omega^{k+1}, \ee
where%
 \be%
  \omega=fdx^{\mu_1}\wedge\cdots\wedge dx^{\mu_k}\in
\Omega^k.%
 \ee%
  When $k=0$, $\Omega^0=\cal A$, then $d:
\Omega^0\rightarrow \Omega^{1}$ is given by%
 \be%
d_Df=\sum_{\alpha}\Delta_\alpha f dx^\alpha.%
\ee%

  It is straightforward to
prove that
\bes\nonumber
(a):&& (d_Df)(v)=v(f), \quad v\in T({ N}), f \in \Omega^0,\\%
(b): &&d_D(\omega\otimes\omega')=d_D\omega\otimes\omega'
+(-1)^{deg\omega}\omega\otimes d_D\omega', \quad\omega, \omega'\in
\Omega^*,\\\nonumber%
 (c): &&d_D^{2}=0,
\ees
provided that the following conditions hold%
 \bes\nonumber
(1) &&  dx^\mu\wedge
dx^\nu=- dx^\nu\wedge dx^\mu ,  \\
(2) &&  d_D(dx^\mu)=0,  \\\nonumber
 (3) &&  dx^\mu f=(E_{\mu}f) dx^\mu,
\   \ \rm (no~ summation).%
 \ees
 Thus, we  set up the well defined differential
algebra. Note that in order to do so the multiplication of
functions and one-forms must be noncommutative.

\section{Difference Discrete Connection and Curvature}\label{DDC}

 \hskip6mm The discrete analogue of connection has been given by
two of the present authors \cite{GWZ} and  others (see, e.g.
\cite{Dim94}, \cite{Dim941}, \cite{Dim942}, \cite{Dim943},
\cite{Nov}, \cite{LMW}).  In this section
 we  define the (difference) discrete connection in simple way similar to that in the continuous
case
  based on the noncommutative differential calculus introduced in the last
  section.
  As was mentioned, the key point  
  is
  to replace the difference discrete exterior derivative  by
the  covariant difference discrete derivative for
  the sections on bundle. \omits{or to introduce the minimal coupling of
  fields with the (discrete) gauge potential in terms of physical
  terminology.}

\subsection{Connection, Gauge Transformations and Holonomy}

\subsubsection{Discrete Bundle and Discrete Sections} \hskip6mm
Let $P=P(\mathcal M, G)$ be a principal bundle  over an
$n$-dimensional base manifold $\cal M$ isomorphic to $R^n$ with a
Lie group $G$ as the structure group. Now consider its discrete
counterpart in the following manner  as a discrete principle
bundle. Let the manifold $\mathcal M$ \omits{locally} be
discretized as $\mathbb{Z}^n$, i.e. the hypercubic lattice with
equal spacing $h=1$, and take such an ${\cal M}\simeq
\mathbb{Z}^n$ (a lattice $N$) as the discrete base space.
\omits{while the fibre as a structure group is still isomorphic to
the Lie group $G$. At a point $x$ as a node on the discretized
base manifold, ${\cal N}_x$ be the set of nodes neighboring to
$x$.
 Let ${I}_x $ the index set of nodes of ${\cal N}_x$ including
 $x$. The union of all ${\cal N}_x$ \omits{will be}is denoted as $N$, which
 is also \omits{the} all nodes in the discrete base manifold.}
 For a node
$x\in  N$, there is a fiber $G_x=\pi^{-1}(x)$ isomorphic to the
Lie group $G$ as the structure group. The union of all these
fibers are called discrete principal bundle and
denoted as $Q( N, G)$:%
 \be%
  Q=\bigcup_{x\in
 N}\pi^{-1}(x).%
 \ee

If the structure group $G$ is a linear matrix group, for example,
$GL(m, R)$ or its subgroup, \omits{then} we may also define an
associated vector bundle $V=V( N, R^m, GL(m,R))$ \omits{with}to
the discrete principal bundle $Q(N, GL(m, R))$. For any $x\in  N$
there is a fiber $F_x=\pi^{-1}(x)$ isomorphic to an
$m$-dimensional linear space $R^m$ with the right action of
$GL(m,R)$ on the $F_x$. The union of all these fibers
\omits{are}is called a discrete vector bundle:
 \be%
  V=\bigcup_{x\in
N}\pi^{-1}(x).%
\ee

 The discrete field on  discrete
 bundle is a section
 $h(x)\in G $  on discrete principal bundle or a
 $\psi^j(x)\in R^m $ on the vector bundle for all $x\in  \mathcal N_x$. A section
 on discrete bundle is the map,
     \be%
      G:  N=\bigcup \mathcal N_x \longrightarrow Q=Q(N, G), \quad G: x\mapsto G(x).%
 \ee%
 Similarly, we can define the discrete section
 on a
 discrete
 vector bundle. The union of the all  sections  is denoted as
 $\Gamma (Q)$ for the discrete principal bundle or  as $\Gamma(V)$ for
 the discrete vector bundle, respectively.

 One of the examples for the discrete vector bundle is the tangent
 bundle $TN$ over an $n$-dimensional hypercubic lattice $N$ in eq.(\ref{TTdual}).
  The base manifold of this bundle is the hypercubic
 lattice, the fiber $\pi^{-1}(x)$ over $x\in N$ is \omits{the}an
 $n$-dimensional vector space with basis $\{\Delta_\mu,
 \mu=1,\cdots, n\}.$ Another example is its dual bundle $T^*N$,
 its fiber is also the $n$-dimensional vector space with basis $\{dx^\mu,
 \mu=1,\cdots, n\}.$

 For the discrete vector bundle the section space $\Gamma(V)$ is
 also the vector space. The same for the $\Gamma(T^*N)$. Their tensor
 product is given by
 \be\label{tp}%
  \Gamma( T^*N\times V)=\Gamma(T^*N)\times \Gamma(V).%
  \ee}
As was mentioned in previous section, the one form basis $dx^\mu$
in discrete case is defined on a point $x\in N$ but links the
point with another nearby point 
$x+\hat{\mu}$. Then the section of one forms in $\Gamma(T^*N)$ is
generally defined on ${\cal N}_x $, its structure is very
different from  the section in continuous case. So we call it as
discrete section. The section in $\Gamma(V)$ may be the discrete
section as in $\Gamma(TN)$ of tangent vector bundle $TN$. However
there is another possibility. Namely, the section is defined on
the node only  as in the discrete principal $G$-bundle.  The
direct product here can be simply understood as the discrete
counterpart of the direct product in the continuous case in the
above manner.

\subsubsection{Definition of Connection and Gauge
Transformations}

{\bf Definition }:  A difference discrete connection or covariant
difference discrete derivative
 is the linear map
 \be
\mathcal D: \Gamma(V)\longrightarrow \Gamma( T^* N \times V),
 \ee
 which  satisfies the following condition:
 \bes\nonumber
  \mathcal
D(s_1+s_2)=\mathcal Ds_1+\mathcal Ds_2,\\
 \label{GisCon}
  \mathcal
D(as)=d_Da\otimes s+a\mathcal Ds,
\ees %
where $s, s_1, s_2 \in \Gamma(V)$ and $a\in \mathcal A$. This is
almost the same as the definition of the connection or the covariant
derivative in the continuous case, which is basically equivalent to
the connection a la Cartan.

 Since our  discussion on all geometric quantities are in the
local sense, for the simplicity, we can choose the basis
$\{s_\alpha,~1\leq \alpha\leq m\}$  as the basis of the linear
space $\Gamma(V)$ and   $\{ dx^\mu\otimes s_\alpha,\quad 1\leq
\alpha\leq m, ~1\leq \mu\leq n\}$ as the basis  of the sections
 space
 $\Gamma(T^* N\times V)$.  Then the covariant derivative $\mathcal
 Ds_\alpha$ should be the linear expansion on $\{ dx^\mu\otimes
 s_\alpha\}.$
 Hence, we can define it in  the local sense as
 \be%
  \mathcal
Ds_\beta=\sum_{\alpha}\big[-(B)^\alpha_\beta\otimes
s_\alpha\big]=\sum_{\alpha,\mu}\big[-(B_\mu
)^\alpha_{\beta}dx^\mu\otimes s_\alpha \big]%
\ee%
 or simply 
\be%
 \mathcal Ds=- B\cdot s,%
 \ee%
 where $B=B_\mu dx^\mu$  is the local expression of discrete connection $1$-form. For  the
continuous case, the connection 1-form is valued on a Lie algebra.
 However the $1$-form $B=B_\mu dx^\mu$ here is matrix
valued. Since all  $1$-forms are defined on the links,  the
coefficients $B_\mu$  are also defined on the link
$(x,x+\hat{\mu})$
and can be written as %
\be%
 B_\mu(x)=B(x,x+\hat{\mu}).%
  \ee%

 For  any
section $S=\sum_\alpha a^\alpha s_{\alpha}$,   we have \be
\ba{rcl} & \mathcal DS&=\dsum(d_Da^\alpha\cdot
s_\alpha+a^\alpha\cdot \mathcal D
    s_\alpha)\\[3mm]
&&=\dsum (\triangle_\mu a^\alpha dx^\mu\otimes s_\alpha
-a^\alpha(B_\mu)_\alpha^\beta
dx^\mu\otimes s_\beta)\\[3mm]
&&=\dsum (\triangle_\mu a^\beta  -a^\alpha(B_\mu)_\alpha^\beta) dx^\mu\otimes s_\beta\\[3mm]
&&=\dsum (\mathcal D_{D\mu} a^\beta) dx^\mu\otimes s_\beta, \ea
\ee
 where%
  \be\label{cd}%
   \mathcal D_{D\mu} a^\beta=\triangle_\mu a^\beta
-a^\alpha(B_\mu)_\alpha^\beta%
\ee%
  is called the discrete  covariant
derivative of the vector $a^\alpha$ and %
\be%
\mathcal D_{D}
a^\alpha=\dsum \mathcal D_{D\mu} a^\alpha dx^\mu=\dsum
(\triangle_\mu a^\alpha -a^\beta(B_\mu)^\alpha_\beta) dx^\mu%
\ee%
 is the discrete exterior covariant derivative of the vector $a^\alpha$.

On the space $\Gamma(V)$, we can choose another linear basis or
perform \omits{locally} the linear  transformation of the basis,
i.e., take\omits{ing} the gauge transformation as follows
 \be\label{gauge}
  s_\alpha\longmapsto
\tilde{s}_\alpha=g(x)^\beta_\alpha \cdot s_\beta, \qquad
x\in N, %
\ee%
where $g(x)^\beta_\alpha$ is the $GL(m, R)$ gauge group valued
function  defined on the node $x\in N$. \omits{Then} In order to
keep the section $S$ being invariant, the coefficients $a^\alpha$ of
a general section
$S=\sum a^{\alpha}s_{\alpha}$ \omits{will}should transform as%
 \be\label{trnsfma}%
  a^\alpha\longmapsto
\tilde{a}^\alpha=a^\beta\cdot (g^{-1}(x))^\alpha_\beta.%
\ee%

From the gauge invariance of $S$ and $\mathcal D S$, we can derive
the gauge transformation property of $\mathcal D_{D\mu}
a^\alpha$,%
 \be\ba{rcl}%
  \mathcal D S\longmapsto& \dsum (\mathcal D_{D\mu}
a^\gamma) dx^\mu\cdot (g^{-1}(x))^\alpha_\gamma\otimes
g(x)^\beta_\alpha\cdot s_\beta\\[1mm]%
&=\dsum (\mathcal D_{D\mu} a^\gamma) \cdot
(g^{-1}(x+\hat{\mu}))^\alpha_\gamma dx^\mu\otimes
g(x)^\beta_\alpha\cdot s_\beta, %
 \ea\ee%
  where the noncommutative commutation relation between function and
 $1$-form basis \omits{are}is used. 
 Then the covariance of the covariant derivative  follows
\be%
  \mathcal D_{D\mu} a^\alpha \longmapsto \mathcal D_{D\mu}
a^\gamma \cdot (g^{-1}(x+\hat{\mu}))^\alpha_\gamma .%
\ee%
Thus, we get the gauge transformation property of the difference
discrete connection $1$-form as%
 \be\label{gauge1} %
 B_\mu(x)dx^\mu\longmapsto
g(x)\cdot B_\mu(x)dx^\mu \cdot g^{-1}(x)  + g(x)\cdot d_D g^{-1}(x),%
\ee%
or %
\be\label{gauge2}%
 B_\mu(x)\longmapsto g(x)\cdot B_\mu(x) \cdot
g^{-1}(x+\hat{\mu}){ +} g(x)\cdot
\triangle_\mu g^{-1}(x).%
\ee%
Together with the gauge transformation property of the
coefficients of a vector field in (\ref{trnsfma}), the gauge
covariance of the derivative $ \mathcal D_{D\mu} a^\gamma$ is
confirmed.

\subsubsection{Discrete Connection via Horizontal Tangent
Vector}\label{sec DCvHV}\hskip 6mm

For the vector bundle, there is another definition of connection. It
is based on the decomposition of the total tangent vector of the
bundle into the horizontal  and vertical parts. Then the horizontal
tangent vector invariant under the right operation of the structure
group also defines a connection. In fact, the horizontal tangent
vector is nothing but the covariant derivatives. This definition can
also be given in an analogous manner  for the discrete case here.

Let us consider the discrete vector bundle  over a discrete base
manifold $ N$ as the regular lattice with the fiber as a smooth
enough $m$-dimensional vector space $V_x$ at $x\in  N$, like the
$V_k$ used in section \ref{DDM}. As we discussed before,
the basis of tangent space $TV_x$  is %
\be\label{basis}%
X_\alpha=\frac{\partial}{\partial a^{\alpha}}, \quad (1\leq
\alpha\leq m),%
 \ee%
where  $a^{\alpha},~1\leq \alpha\leq m,$ are the coordinates of the
fiber $V_x$. The basis of tangent vector on discrete regular lattice
is \be\label{basis1} \triangle_\mu, \quad
(1\leq \mu\leq n).%
\ee%
 Then the basis of total tangent space
of discrete vector bundle is the union of (\ref{basis}) and
(\ref{basis1}).

Similar to the continuous case, the vector space tangent to the
fiber, i.e. the linear combination of basis in (\ref{basis}), is a
vertical subspace of the total tangent space of the discrete vector
bundle. Its complementary vectors of the vertical subspace in the
total tangent of vector bundle are horizontal and constitute the
horizontal subspace. The basis of horizontal
subspace is as follows,%
 \be \label{basis2}%
  X_\mu=\triangle_\mu-
(B_\mu)^\alpha_\beta a^\beta\frac{\partial}{\partial a^{\alpha}},
\quad (1\leq \mu\leq n).%
\ee%
In comparison with the definition of difference discrete
connection, it is easy to see that the horizontal vector is
nothing but the covariant derivative in  (\ref{cd}).

This means that form the decomposition of tangent vector  on the
total bundle space  we get the coefficients $B_\mu(x)$ of the
discrete connection.  For a given
difference discrete connection or its coefficients $B_\mu(x)$, we can also  get a 
decomposition of the total tangent vector space of bundle into
vertical and horizontal parts sufficiently and necessarily. This
shows that the difference discrete connection on discrete vector
bundle is equivalent to  a decomposition of the total tangent
vector space into vertical and horizontal subspaces as above.

  Similarly, we can define the basis of  dual space for the
  decomposition, i.e., the basis of the vertical and horizontal $1$-form space, respectively, as
\be%
 \omega^\alpha=da^\alpha+(B_\mu)^\alpha_\beta a^\beta
dx^\mu,\quad (1\leq \alpha\leq m),%
\ee%
 and%
\be%
\omega^\mu=dx^\mu, \quad (1\leq \mu\leq n).%
\ee%
 They satisfy the
following dual relation:%
\bd%
\omega^\alpha(X_\beta)=\delta^\alpha_\beta, \quad
\omega^\alpha(X_\mu)=0,\quad \omega^\mu(X_\beta)=0, \quad
\omega^\mu(X_\nu)=\delta^\mu_\nu.%
\ed

\subsubsection{Covariant Derivative and Parallel Transport}
\hskip6mm From the above discussions, we can get the difference
discrete connection on discrete vector bundle through the
definition of the absolute derivative or the horizontal tangent
vector. Both lead to  the  difference discrete covariant
derivative for the vectors,
\be%
 {\cal D}_{D\mu} a^\beta=\triangle_\mu
a^\beta -a^\alpha(B_\mu)_\alpha^\beta. \ee In terms of the relation
\omits{of}between $\triangle_\mu$ and $E_\mu$,  we obtain another
expression for covariant derivative as,
 \be%
 \mathcal D_{D\mu} a^\gamma(x)=E_\mu
a^\gamma(x)-a^\beta(x) \cdot [(B_\mu(x))^\gamma_\beta +
\delta^\gamma_\beta],%
\ee%
 where $\delta^\alpha_\beta$ is the unit
matrix.   Then we obtain another expression for the coefficient of
discrete connection,
 \be%
  U_\mu(x)=[(B_\mu(x))^\gamma_\beta +
\delta^\gamma_\beta],\ee
 which  is an element of some {group}, for example the group
 $GL(m, R)$ in our discussions, and connects the
points $ x$ and $x+\hat{\mu}$. We can also call $U_\mu(x)$ as the
discrete $GL(m, R)$-connection and express it as
\be\label{52}%
  U_\mu(x)=U(x,x+\hat{\mu}).\ee
From the second expression of the coefficient of connection and
the definition of covariant derivative for the vectors, it follows 
the parallel transport of the section of vector $a^\beta$ if  its
covariant derivative is zero
 \be%
  {\cal D}_{D\mu} a^\beta=\triangle_\mu a^\beta
-a^\alpha(B_\mu)_\alpha^\beta=0\ee or \be E_\mu
a^\gamma(x)-a^\beta(x) \cdot (U_\mu(x))^\gamma_\beta=0.%
\ee%
Namely,%
 \be%
  a^\gamma(x+\hat{\mu})=a^\beta(x) \cdot (U(x,
x+\hat{\mu}))^\gamma_\beta.
\ee%
 This means that the parallel
transport of the section of vector $a^\beta(x)$ along the path
$x\mapsto x+\hat{\mu}$ is expressed as
\be\label{para}%
 a^\beta(x)\longmapsto
a^\beta(x+\hat{\mu})=a^\gamma(x) \cdot
(U(x,x+\hat{\mu}))_\gamma^\beta.%
\ee

It is shown that there is a parallel transport on the discrete
bundle along the curve on discrete base manifold for a given
discrete connection on discrete bundle.

 Due to  the 
 discrete
connection coefficient $U(x,x+\hat{\mu})$  as a   group element,
there are the following group properties of $U(x,x+\hat{\mu})$
along the path decomposition of $(x,x+\hat{\mu}+\hat{\nu})$ into
$(x,x+\hat{\mu})$
 and $(x+\hat{\mu}, x+\hat{\mu}+\hat{\nu})$,
 \be%
  U(x,x+\hat{\mu})\cdot U(x+\hat{\mu},
x+\hat{\mu}+\hat{\nu})=U(x, x+\hat{\mu}+\hat{\nu}).\ee
 From the inverse of the parallel
transport, it also follows that
\be%
(U(x,x+\hat{\mu}))^{-1}=U(x+\hat{\mu}, x).%
\ee

 The coefficient of the discrete connection $U(x,x+\hat{\mu})$
determines the parallel  transport not only on a vector bundle but
also on a $GL(m, R)$ principal bundle. Therefore, it is also
called a discrete $GL(m, R)$ connection.

\subsubsection{Parallel Transport on Discrete Principal Bundle}
\hskip6mm
 The  matrix structure 
 group  of discrete
vector bundle can be generalized to any Lie group $G$ and the
coefficients of a connection are the group $G$-valued  with the
right \omits{gauge transformations}operations. Thus, we can  get a
difference discrete $G$-valued connection on a discrete principal
$G$-bundle over the lattice.

In this case,  the concept of parallel transport can be extended
to the discrete principal $G$-bundle. For a section $h(x)$, the
parallel transport with respect to a $G$-valued connection
$U(x,x+\hat{\mu})$ reads
\be%
 h(x)\longmapsto h(x+\hat{\mu})=h(x) \cdot
U(x,x+\hat{\mu}).%
\ee%
All elements here  belong to the Lie group $G$ and multiplication
should be the group multiplication. The above equation can be
expressed as %
\be%
 h(x_0)\longmapsto h(x_1)=h(x_0) \cdot U(x_0,x_1),%
 \ee%
or %
\be%
 h(x_1)\longmapsto h(x_0)=h(x_1) \cdot U(x_1,x_0),%
 \ee%
 which
 implies that%
 \be%
  U(x_1, x_0)=U(x_0, x_1)^{-1}.%
  \ee%

Similarly, the covariant derivative for the 
section $h(x)$ can be given as
\be%
 \mathcal D_{D\mu}h(x)= E_\mu h(x)-h(x) \cdot
U_\mu(x),%
\ee%
 and the covariant exterior derivative as%
 \be%
  \mathcal
D_{D}h(x)=\mathcal D_{D\mu}h(x)dx^\mu=\sum_\mu (E_\mu h(x)-h(x)
\cdot U_\mu(x))dx^\mu.%
\ee%

\omits{\subsection{Holonomy group} 
Now we  discuss the
holonomy group on the difference discrete principal bundle
$Q(\mathcal N, G)$. If there is a discrete curve
$\gamma=\{x_0,x_1,\cdots, x_{n-1},x_n=x_0\}$ on discrete base
manifold $\mathcal N$ and  a point $h_0\in \pi^{-1}(x_0)$, the
corresponding  parallel transport of $(x_0, h_0)$ along the curve
$\gamma$ is given by the formula \be h(x_{j+1})=h(x_{j})\cdot
U(x_j,
x_{j+1}),\qquad j=0,1,\cdots,n-1.%
\ee%
Then one should have%
 \be%
  h(x_n)=h(x_0)\prod_{j=0}^{n-1}U(x_j,
x_{j+1}).%
\ee
 When  $h_0=h_n$,  the above equation leads to
 \be%
  \prod_{i=0}^{n-1}U(x_i, x_{i+1})=1,
  \ee%
   which
 is corresponding to the $0$-curvature. Otherwise, the curvature is
 nontrivial or the holonomy is not trivial.}

\subsection{Difference Discrete Curvature, Bianchi Identity and Abelian Chern Class}

\subsubsection{Difference Discrete Curvature}\hskip6mm Based on the definition of the difference
discrete connection $1$-forms on discrete vector bundle, we can
define the curvatures $2$-forms similar to the continuous case
\cite{DFNC1}, \cite{DFNC2}
 \be%
  F=d_DB+B\wedge B.%
  \ee%
   If we assume $F=\dfrac
12F_{\mu\nu}dx^\mu\wedge dx^\nu$, it follows that %
 \be\label{Fmunu}%
F_{\mu\nu}(x)=\triangle_\mu B_\nu(x)-\triangle_\nu
B_\mu(x)+B_\mu(x)\cdot B_\nu(x+\hat{\mu})-B_\nu(x)\cdot
B_\mu(x+\hat{\nu}).%
\ee%
Under the continuous limit, it is easy to see that this curvature
should be the same as the usual formula of curvature.

Since the definition is in a similar formulation except for the
non-commutative exterior differential calculus, it is also easy to
check the covariance property of the curvature under the gauge
transformation (\ref{gauge1}) or (\ref{gauge2}) as follows
 \be F(x)\longmapsto
\widetilde{F}(x)=g(x)\cdot F(x) \cdot g^{-1}(x),\ee or the
covariance behavior of its components%
 \be%
  F_{\mu\nu}(x)\longmapsto
\widetilde{F}_{\mu\nu}(x)=g(x)\cdot F_{\mu\nu}(x) \cdot
g^{-1}(x+\hat{\mu}+\hat{\nu}).%
 \ee%

It is important to see that from the non-commutative property of
differential calculus on lattice the shifting operator appears in
the covariance of discrete curvature. This is a main difference
between the continuous case and the discrete one. And it may lead
to more difficulties in discussion of the gauge covariance and
invariance property of tensors in the discrete case.

\subsubsection{Curvature via Holonomy}\hskip6mm In continuous case,
the curvature may naturally appear in  the homolomy consideration.
\omits{Generally speaking, there is no unique way to define the
curvature.} As the (difference) discrete counterpart, the
(difference) discrete curvature may also be described based on the
holonomy consideration. \omits{Now we try to define the curvature
in a similar way for the continuous case.}

 Let us consider the squire
of the exterior covariant derivatives
\be\ba{rcl}%
 && (\mathcal D_{D})^2 h(x)\\[2mm]&=&\mathcal
D_{D} \mathcal D_{D\mu}h(x)dx^\mu\\[2mm]%
&=&\mathcal D_{D}\dsum_\mu
(E_\mu h(x)-h(x) \cdot U_\mu(x))dx^\mu\\[2mm]%
&= &\dsum_{\mu\nu}
E_\nu(E_\mu h(x)-h(x) \cdot U_\mu(x))dx^\mu\wedge
dx^\nu\\[2mm]%
&&-\dsum_{\mu\nu} (E_\mu h(x)-h(x)\cdot U_\mu(x))dx^\mu
\wedge U_\nu(x))dx^\nu\\[2mm]%
&=&h(x)\dsum_{\mu\nu} U_\mu(x)dx^\mu
\wedge U_\nu(x)dx^\nu\\[2mm]%
&=&h(x)\dsum_{\mu\nu}
U_\mu(x)\cdot U_\nu(x+\hat{\mu})dx^\mu \wedge dx^\nu\\[2mm]%
&=&\dfrac 12h(x)\dsum_{\mu\nu} [U_\mu(x)\cdot
U_\nu(x+\hat{\mu})-U_\nu(x)\cdot U_\mu(x+\hat{\nu})]dx^\mu \wedge
dx^\nu.%
\ea\ee%
This leads to another expression for the curvature $2$-form with
its
coefficients 
\be\ba{rcl}%
 &&F=U^2=\dsum_{\mu\nu} U_\mu(x)dx^\mu
\wedge U_\nu(x)dx^\nu,\\ &&F_{\mu\nu}=\dfrac 12[U_\mu(x)\cdot
U_\nu(x+\hat{\mu})-U_\nu(x)\cdot U_\mu(x+\hat{\nu})].%
\ea\ee%
  The zero curvature
condition $F=0$ is%
 \be\label{4U}
   U_\mu(x)\cdot U_\nu(x+\hat{\mu})
=U_\nu(x)\cdot U_\mu(x+\hat{\nu}), \ee or\be U_\mu(x)\cdot
U_\nu(x+\hat{\mu}) \cdot U^{-1}_\mu(x+\hat{\nu})\cdot
U^{-1}_\nu(x)=1,\ee or \be\label{54} U(x,x+\hat{\mu})\cdot
U(x+\hat{\mu},x+\hat{\mu}+\hat{\nu}) \cdot
U(x+\hat{\mu}+\hat{\nu},x+\hat{\nu})\cdot U(x+\hat{\nu},x)=1.%
\ee%
 The expressions in last formula is nothing but the holonomy group
in geometry or the  plaquette  variable in lattice gauge theory.
We will  discuss them in the  section 6 and  show also  that zero
curvature condition is just the integrable condition for a
discrete integrable system.

\subsubsection{Bianchi Identity and Abelian Chern Class}
\hskip6mm Similar to the Bianchi identity in differential geometry,
we can also derive the Bianchi
identity for the difference discrete curvature
\be%
 {\mathcal D_D}F=d_DF-F\wedge
B+B\wedge F=0,%
\ee%
 or in its components%
  \be%
\varepsilon^{\lambda\mu\nu}[\triangle_\lambda
F_{\mu\nu}(x)-F_{\lambda\mu}(x)\wedge
B_\nu(x+\hat{\mu}+\hat{\nu})+B_\lambda(x)\wedge
F_{\mu\nu}(x+\hat{\lambda})]=0.%
\ee%

For the Abelian case,  one can define the following topological term
as discrete Chern class \cite{FSW}, \cite{FSW1}, \cite{FSW2} ,
   \be%
    c_k=F\wedge F\wedge \cdots \wedge F,%
    \ee%
which was used to discuss the chiral anomaly in the lattice gauge
theory. The coefficient of the Abelian Chern class is %
\bd%
\varepsilon^{\mu_1\mu_2\cdots\mu_{2k-1}\mu_{2k}}F_{\mu_1\mu_2}(x)\cdot
F_{\mu_3\mu_4}(x+\hat{\mu_1}+\hat{\mu_2})\cdots
F_{\mu_{2k-1}\mu_{2k}}(x+\hat{\mu_1}+\hat{\mu_2}+\cdots+\hat{\mu_{2k-2}}).
\ed%
  This equation was first appeared  in lattice gauge
theory for the 
Abelian anomaly of  chiral fermion in a quantum field theory
\cite{FSW},\cite{FSW1}, \cite{FSW2}.

\section{Discrete Connection on $G$-Bundle
over Random Lattice}\label{sec:DConRL}

\hskip6mm The definition of the discrete connection via the
horizontal vector space 
in sect. \ref{sec DCvHV} can be generalized to the one on a
$G$-bundle $Q( N, G)$ over a random lattice $ N$.

Let us consider the parallel transport of a section on such a
$G$-bundle:
 \be\label{h-para}%
h(x_0)\longmapsto h(x_1)=h(x_0) \cdot U(x_0,x_1),%
\ee%
 where
$h(x_j)$ is the $G$-valued section defined on $x_j$, $j=0,1$ and
$x_0$, $x_1$ are nearest
neighbor. We can reexpress equivalently it as
 \be%
  (x_0,
h_0)\longmapsto (x_1, h_1)=(x_0, h_0) \cdot U(x_0,x_1),%
\ee%
 where $h_0=h(x_0), ~h_1=h(x_1)$ and
right multiplication of $U$ on the bundle  acts only  on the
$G$-valued section $h_0$. It is easy to see that in these
expressions there is no difference operator involved so that they
could be make
sense for the $G$-bundle over random lattice, if the discrete
connection is properly introduced. 

On the other hand, if $h_0$ and $h_1$ satisfy eq.(\ref{h-para}),
it can be proved that \omits{according to the result of Leok et al
\cite{LMW}, }the element $(q_0, q_1)=((x_0, h_0),(x_1, h_1))\in
Q\times Q$ is just a horizontal vector on
$TQ$, i.e.
 \be%
  \mbox {hor}((x_0, h_0),(x_1, h_1))=((x_0,
h_0),(x_1, h_1)),%
\ee%
  where $\rm hor(*,*)$ denotes the horizontal part of the $(*,*)$.
In fact, this is almost the same as the one introduced in
\cite{LMW}.

Thus, our definition for the discrete connection can be easily
compared with the local expression $A(x_0, x_1)$ of the
coefficients
of a connection $1$-form defined in \cite{LMW}. Namely, %
 \be%
U(x_0,x_1)=A(x_0, x_1)^{-1}%
\ee%
 then%
  \bd ((x_0,h_0), (x_1, h_0\cdot
A(x_0, x_1)^{-1}))%
 \ed%
  is a horizontal vector. According to the
formulation in \cite{LMW}, we get%
 \be\ba{rcl}%
  &&((x_0,h_0), (x_1,
h_0\cdot  A(x_0, x_1)^{-1}))\\[2mm]%
&&=h_0\cdot  i_{(x_0,e)}( A(x_0, x_1)^{-1})\cdot ((x_0,e), (x_1,
e))\\[1mm]%
&&=h_0\cdot \mbox{hor}((x_0, e), (x_1,e))\\[1mm]
&&=h_0\cdot \mbox{hor}((x_0, e), q_1)\\[1mm]
&&=h_0\cdot \mbox{hor}((x_0, e),
h^{-1}_0q_1)\\[1mm]
&&=\mbox{hor}((x_0, h_0), q_1),%
 \ea\ee%
  which means that the horizontal
vector $((x_0,h_0), (x_1, h_0\cdot  A(x_0, x_1)^{-1}))$ is the
horizontal part of any vector $(q_0, q_1)$ with $q_0$ is fixed and
$q_1$ is any point on the fiber of $\pi^{-1}(x_1)$.

According to the definition of $A(x_0, x_1)$,  we have
\[A(x_0,x_1)=\mathcal{A}_d(x_0,e,x_1,e)\, .\] %
From  the gauge transformation property of $U(x_0, x_1)$, it
follows that
under the gauge transformation $g(x)$%
 \be%
  A(x_0, x_1)\longmapsto
g(x_1)\cdot A(x_0, x_1)\cdot g^{-1}(x_0).%
 \ee%
This leads to 
\be%
\mathcal{A}_d(x_0,g(x_0);x_1,g(x_1))=g(x_1)\mathcal{A}_d(x_0,e,x_1,e)g^{-1}(x_0).%
\ee
Thus, we recover the  property of the connection $1$-form defined
 in \cite{LMW}
 \be
\mathcal{A}_d(gq_0,hq_1)=h\mathcal{A}_d(q_0,q_1)g^{-1}.%
\ee%

 As was shown above, our
definition of discrete connection is equivalent to that in
\cite{LMW} in the case of the cubic  lattice. However, our
definition  for the discrete curvature is only for the hypercubic
lattice, since it is based on the noncommutative differential
calculus. How to extend those results to the case of random
lattice is under investigation.

\section{Applications }\label{sec appl}

\subsection{Lattice Gauge Theory and Difference Discrete
Connection}\label{sec lgt}

\hskip6mm In the lattice gauge theory \cite{Ro}, the space-time is
discretized as hypercubic lattice with equal spacing $a$ in any
direction  in most cases.

Suppose that $A_\mu$ is the gauge field or the connection on the
continuous case. At each link on the lattice we introduce a
discrete version of
the path ordered product%
\be%
 U(x,x+\hat{\mu})\equiv
U_\mu(x)=\disp e^{ ia A_\mu(x+\frac {\hat{\mu}}2)}, %
\ee %
 where $\hat{\mu}$ is the vector in coordinate direction with
length $a$ and $x$ is the point coordinates on the node\omits{note}
of the hypercubic lattice which takes integer value only. The
average field, which is denoted by $A_\mu(x+\frac {\hat{\mu}}2)$, is
defined at the midpoint of the link $(x, x+\hat{\mu})$.
Similarly, %
\be%
 U(x,x-\hat{\mu})\equiv U_{-\mu}(x)=\disp e^{- ia
A_\mu(x-\frac {\hat{\mu}}2)}=U^\dag(x-\hat{\mu},x).%
\ee%
 If the
connection $A_\mu$ is  \omits{ taking} valued on the Lie algebra of
$SU(N)$ with a hermitian basis, \omits{then} we have%
 \bd%
U^\dag(x-\hat{\mu},x)=U^{-1}(x-\hat{\mu},x).%
\ed%

The variable of a simplest  Wilson loop  called  plaquette variable
is expressed as the left side of (\ref{54}),
which is defined on the two dimensional square%
 \be%
W_{\mu\nu}=U_\mu(x)\cdot U_\nu(x+\hat{\mu})\cdot
U_\mu^\dag(x+\hat{\nu})\cdot U_\nu^\dag(x).%
 \ee

It  can be shown that the continue limit of $W_{\mu\nu}$
\omits{will}
is related to the Yang-Mills action%
 \bd %
 \mbox
{Re}(1-W_{\mu\nu})=\dfrac{a^4}2F_{\mu\nu}F^{\mu\nu}+O(a^6)+\cdots,%
\ed%
\bd%
 \mbox {Im}(W_{\mu\nu})=a^2F_{\mu\nu}+\cdots. %
 \ed
 Therefore, plaquette variable $W_{\mu\nu}$ should play some rule of curvature
in the discrete case as we discussed in previous section. However
its continuous limit is related  not only to the usual curvature but
also  to the Yang-Mills action as in the above expressions, there
should be more geometric meaning in the theory of discrete
connection and curvature than usual one.

\subsection{Geometric Meaning of Discrete Lax Pair }\label{sec lax}\hskip6mm
In order to understand the discrete connection on discrete bundle,
 we first discuss some geometric meanings of the Lax pair
and discrete Lax pair. In fact,  it gives one of  solid
motivations and some consideration  for the study of discrete
connection.

 Let us start with  the concept of  Lax pair of integral
system in continuous  two dimensions case  with $1$-dimension time
and $1$-dimension space as follows%
 \be\ba{rcl}\label{15}
&&\partial_x\psi=\psi\cdot A_x,\\[1mm]
&&\partial_t\psi=\psi\cdot A_t,%
\ea\ee%
 where $\psi$ is a vector and
$A_x, A_t$ are matrix valued. The consistent condition for this
linear system is \omits{\be
\partial_x\partial_t\psi=\partial_t\partial_x\psi \ee for any
$\psi$, it is equivalent to \bd \psi\cdot \partial_x A_t+\psi\cdot
A_x\cdot A_t=\psi\cdot\partial_t A_x+\psi\cdot A_t\cdot A_x \ed
i.e.,}
\be\label{17}
\partial_xA_t-\partial_tA_x+[A_x, A_t]=0. \ee

\omits{When we call (\ref{15}) as the covariant derivative, \bd
D_x\psi=0, \qquad D_t\psi=0,\ed then equation (\ref{17}) is
nothing but the zero curvature condition for connections $A_x,
A_t$, i.e., \bd F_{xt}=\partial_xA_t-\partial_tA_x+[A_x, A_t].\ed}

 Now let us discretize the $2$-dimensional space-time as a square
lattice, $R^2\rightarrow \mathbb{Z}^2$. The section field
$\psi(x)$ on the vector bundle becomes the field $\psi(m,n)$ the
functions depending on two discrete variable, i.e., two integer
$(m,n)$. Naively the discrete Lax pair may be written as
\be\ba{rcl}\label{18}%
 &&\triangle_x \psi(m,n)=\psi(m,n)\cdot A_x(m,n),\\[1mm]
&&\triangle_t \psi(m,n)=\psi(m,n)\cdot A_t(m,n),%
\ea\ee %
The derivatives $\partial_x$ and $\partial_t$ with respect to $x$
and $t$ are replaced by difference operators $\triangle_x$ and
$\triangle_t$, respectively. \omits{defined as follows,
\bd\ba{rcl} &&\triangle_x\psi(m,n)=\dfrac
1{\triangle x} (\psi(m+1,n)-\psi(m,n))\\[3mm]
&&\triangle_t\psi(m,n)=\dfrac 1{\triangle t}(\psi(m,
n+1)-\psi(m,n))\ea\ed where $\triangle x$ and $\triangle t$ are
lattice spacing length in $x$ and $t$ direction, for the
simplicity we take both of them as $1$.}
 The  consistent condition
for the discrete Lax pair, i.e., \be \triangle_x\triangle_t
\psi(m,n)=\triangle_t\triangle_x \psi(m,n)\ee  leads to
\be\label{24} \triangle_x A_t(m,n)-\triangle_t
A_x(m,n)+A_x(m,n)A_t(m+1,n)-A_t(m,n)A_x(m,n+1)=0. \ee

 Using the shift operator $E_x$
 and $E_t$
\omits{
 defined as
 \bd\ba{rcl} &&E_x\psi(m,n)=\psi(m+1,n),\\[1mm]&&
 E_t\psi(m,n)=\psi(m,n+1),\ea\ed}
we can also rewrite the discrete Lax pair (\ref{18}) as
\be\ba{rcl}\label{dLax} &&E_x\psi(m,n)=\psi(m,n)\cdot
[1+A_x(m,n)]=\psi(m,n)\cdot
U_x(m,n)\\[1mm]
&&E_t\psi(m,n)=\psi(m,n)\cdot [1+A_t(m,n)]=\psi(m,n)\cdot
U_t(m,n),\ea\ee where \be\label{20}  U_t(m,n)=1+A_t(m,n),\qquad
U_x(m,n)=1+A_x(m,n). \ee

 On requiring the corresponding consistent condition \be
E_xE_t\psi(m,n)=E_tE_x\psi(m,n),\ee  a straightforward calculation
leads to \be\label{22} U_x(m,n)\cdot U_t(m+1,n)=U_t(m,n)\cdot
U_x(m,n+1),\ee or
 \be\label{23} U_x(m,n)\cdot
U_t(m+1,n)\cdot U^{-1}_x(m,n+1)\cdot U^{-1}_t(m,n)=1. \ee If we use
the relation of $U$ and $A$ in (\ref{20}), we can derive the zero
curvature condition (\ref{24}) form (\ref{23}).

When we regard the quantities $A_x$, $A_t$, $U_x$ and $U_t$ as the
discrete connections on discrete bundle,  the equations (\ref{24})
and (\ref{23}) should be the zero curvature condition for these
connections, and the left sides of (\ref{24}) and (\ref{23})
should be the extension of the curvature in the discrete case.

\section{Remarks and Discussions}

\hskip6mm {As was mentioned  previously,  the study of discrete
models is very important in both their own right and applications,
although we mainly focus on the discrete models as the 
discrete counterparts of the continuous cases. \omits{Usually, the
 way  of studying these models is 
 structure preserving discretization of the continuous models,
 such that}

 In order to get the discrete models that can
 keep the properties of continuous ones as much as possible, \omits{. In computational
 mathematics there are lot of structure preserving algorithms \cite{FK},
 which  mostly  {\green come from  the}  discretization of original ODE or PDE.

 Another way to get the
 symplectic or multi-symplectic  structure preserving algorithm for
 Hamiltonian mechanics or field theory are the discrete variation
 method. Where in stead of discretizing the original equation,
{\green one }}we may first consider  a kind of discrete  models
from their continuous counterparts with differences as discrete
derivatives. These models can be given by replacing the both
continuous independent variables and their derivatives by the
discrete independent variables as a regular lattice and their
differences on the lattice, respectively. \omits{ then {\green
uses} the discrete variation to get difference equation
 which  {\green is}  corresponding to the symplectic or multi-symplectic  structure preserving
 algorithm.  {\green}}In general, for the discrete models on the regular lattices including the
 models just mentioned, it is natural to study first \omits{gives us a hint  for the study of discrete
 model, i.e.,
 to build it  from
 beginning. First we study }the properties of the function spaces on the lattices and the discrete bundles
 over the lattices both analytically and geometrically,\omits{
  discrete spaces and   their  geometric
 properties,} such as discrete differential calculus, discrete metric,
 discrete Hodge operator, discrete connection and
 curvature, and so on.  In doing so, we may follow a way similar to that in the continuous cases,
 as long as the differences
 are regarded as the discrete derivatives. \omits{to get
 some discrete models or some discrete equations based on
 discrete geometry.   This is the main
 purpose of
 for studying the geometry on the lattice in this paper.}

 In this paper, we have 
briefly reviewed the non-commutative differential calculus on
hypercubic lattice, which have discussed by many groups. We have
mainly introduced the (difference) discrete connections on
discrete vector bundle in several manners, the parallel transport,
the decomposition of the vector space  into vertical and
horizontal space, the covariant
derivative on the section of vector bundle as well as 
the discrete curvature of the discrete connections. We  have also
studied their relation to 
the lattice gauge theory and applied to the Lax pairs for the
discrete integrable systems.

\omits{Most our results on the 
discrete models are  very similar formally to the ones in the
continuous theory.  }There are, of course, also many properties of
the discrete models, which are very different apparently from the
continuous ones. These should be investigated further. Although
 one of the definitions for the discrete connection can be extended to the case over the random lattice,
for the discrete curvature on the random lattice in the lattice
gauge theory, however, it is still open whether it can be defined
in a way similar to the continuous case formally. \omits{we have
given several definitions of curvature but only on the hypercubic
lattice, however we do still not know how to define curvature on
the random lattice. }This is also a very interesting question 
to get more results on lattice with no-trivial topology. Another
very important problem is how to get the topological classes with
non-Abelian group. Needless to say, more attention should be payed
to those questions.


\bigskip

The work is partly supported by NKBRPC(2004CB318000), Beijing
Jiao-Wei Key project (KZ200310028010) and NSF projects (10375087,
10375038, 90403018, 90503002). The authors would like to  thank
Morningside  Center for Mathematics, CAS. Part of the work was
done during the Workshop on Mathematical Physics there.

\end{document}